\documentclass{article}
\usepackage{graphicx} 
\usepackage{hyperref}
\usepackage{amsgen,amsmath,amstext,amsbsy,amsopn,amssymb}
\usepackage{amsmath}
\usepackage{multirow}
\usepackage{xspace}
\usepackage{epstopdf}
\usepackage{authblk}
\usepackage{xcolor}
\usepackage{xr-hyper}
\usepackage{hyperref}
\usepackage{natbib}
\usepackage[english]{babel}
\usepackage{amsthm}
\usepackage{geometry}
\usepackage{algorithm}
\usepackage{algpseudocode}
\usepackage{amsthm}
\usepackage{makecell}
\newtheorem{prop}{Proposition}
\providecommand{\keywords}[1]{\textbf{\textit{Keywords:}} #1}
\title{Confirmatory Biomarker Identification with $k$-FWER Control Using Derandomized Knockoffs with Cox Regression}
\author[1]{Rui Liu}
\author[2]{Nan Sun}

\affil[1]{Department of Statistics and Operations Research,  The University of North Carolina at Chapel Hill, Chapel Hill, North Carolina, USA}
\affil[2]{BeOne Medicines, Shanghai, China}
\date{}

\begin{document}
	
	\maketitle
	\begin{abstract}
		Selecting important features in high-dimensional survival analysis is critical for identifying confirmatory biomarkers while maintaining rigorous error control. In this paper, we propose a derandomized knockoffs procedure for Cox regression that enhances stability in feature selection while maintaining rigorous control over the $k$-familywise error rate ($k$-FWER). By aggregating across multiple randomized knockoff realizations, our approach mitigates the instability commonly observed with conventional knockoffs. Through extensive simulations, we demonstrate that our method consistently outperforms standard knockoffs in both selection power and error control. Moreover, we apply our procedure to a clinical dataset on primary biliary cirrhosis (PBC) to identify key prognostic biomarkers associated with patient survival. The results confirm the superior stability of the derandomized knockoffs method, allowing for a more reliable identification of important clinical variables. Additionally, our approach is applicable to datasets containing both continuous and categorical covariates, broadening its utility in real-world biomedical studies. This framework provides a robust and interpretable solution for high-dimensional survival analysis, making it particularly suitable for applications requiring precise and stable variable selection.
		
	\end{abstract}
	
	
	\section{Introduction}
	
	High-dimensional data have become ubiquitous in biomedical research, driven by advances in high-throughput technologies such as genomic sequencing, transcriptomics, and proteomics. A key challenge in this context is identifying a subset of relevant variables, often referred to as {\it variable selection}, to build interpretable models and distinguish meaningful signals from noise. Variable selection is crucial in various fields, such as identifying biomarkers in genomics, selecting risk factors in clinical studies, and determining relevant predictors in financial and industrial applications. 
	
	Several methods have been developed for variable selection in high-dimensional settings. Among them, penalized regression techniques such as the Lasso \citep{tibshirani1996regression}, Ridge regression \citep{hoerl1970ridge}, stepwise linear regression \citep{hocking1976biometrics}, and their variants have been widely applied in fields dealing with large-scale data, including economics \citep{wang2023variable}, environmental science \citep{li2021radiative}, industrial engineering \citep{sun2021hybrid} and biomedical research \citep{dai2022feature}. These methods offer effective solutions for regularization and prediction but often face challenges when predictors exhibit high correlations, leading to instability and inconsistency in the selected variables. Moreover, Classical variable selection techniques typically lack rigorous guarantees for controlling the number of false discoveries—a significant limitation when testing many variables simultaneously \citep{candes2018panning}.

	To address this issue, researchers have explored various methods for error rate control during variable selection. One commonly used criterion is the false discovery rate (FDR), which controls the expected proportion of false discoveries among the selected variables \citep{benjamini1995controlling}. Controlling FDR in high-dimensional variable selection poses significant challenges due to complex dependencies among variables. The development of the {\it Knockoff filter} by Barber and Candès \citep{barber2015controlling} marked a major breakthrough by providing a flexible and powerful approach for variable selection with rigorous FDR control. This method generates synthetic variables that preserve the dependence structure of the original variables, enabling reliable estimation of variable importance while maintaining control over the false discovery rate. The original knockoff framework has since been extended and generalized, including {\it Model-X knockoffs} \citep{candes2018panning}, which allow for arbitrary models of the covariates, and {\it Deep knockoffs} \citep{Romano2020}, which use deep learning to construct knockoff variables in highly complex and nonlinear settings.
	
	While FDR is widely used in exploratory studies \citep{sesia2019gene,zhu2021deeplink,sechidis2021using, jiang2021knockoff}, it is not the only relevant error criterion. In confirmatory studies, such as clinical biomarker validation, controlling the $k$-familywise error rate ($k$-FWER) is often more appropriate \citep{lehmann2012generalizations,sarkar2009generalized}. The $k$-FWER criterion restricts the probability of making at least \(k\) false discoveries, offering stricter error control compared to the false discovery rate (FDR). This is particularly important in fields where even a small number of false discoveries can lead to serious consequences, such as misinterpreted clinical trial results or misallocated resources in drug development \citep{Janson2016}. By setting \(k = 1\) or \(k = 2\), researchers can limit the number of false positives while still maintaining sufficient power to detect true signals.
	
	Several advancements have been made to adapt knockoff methods for $k$-FWER control. \citep{Janson2016} proposed a method that extends the knockoff framework to $k$-FWER control by using sequential testing procedures. Recent work by  \citep{ren2023derandomizing} on {\it Derandomized knockoffs} addresses a key limitation of traditional knockoffs—their inherent randomness—by aggregating knockoff statistics across multiple random constructions, thus reducing variability and improving reproducibility in variable selection.
	
	Another significant advancement involves extending knockoff-based variable selection to survival analysis. The Cox proportional hazards model \citep{cox1972regression} is a fundamental tool for modeling time-to-event outcomes, and identifying important variables via Cox regression is critical in clinical research \citep{ishwaran2010high,benner2010high,zhang2024robust,wang2024fitting}. Recent studies have adapted model-X knockoffs to a range of survival settings: for the Cox model, \citet{hu2024model} developed a controlled variable-selection approach under proportional hazards with a heterogeneity parameter, and \citet{li2023coxknockoff} established finite-sample FDR control with power guarantees; for accelerated failure time (AFT) models, \citet{dong2022reproducible} and \citet{yu2024reproducible} proposed model-X and deep-knockoff methods with FDR control—the latter accommodating both continuous and categorical covariates; for survival mediation analysis, \citet{tian2022coxmkf} leveraged aggregation of multiple knockoffs; and for the additive hazards model, \citet{dong2025ah} introduced an FDR-controlled approach.  Collectively, these advances demonstrate the suitability of knockoff methodology for high-dimensional survival analysis across multiple hazard formulations.
	
	Specifically, for confirmatory analyses that require stringent error control, controlling $k$-FWER rather than FDR is more appropriate. In this work, we build on these advancements by adapting the derandomized knockoff framework to the Cox proportional hazards model to control the $k$-FWER in variable selection. Our approach leverages the robustness and reproducibility of derandomized knockoffs (\citep{ren2023derandomizing}) while addressing the challenges posed by high-dimensional clinical trial data. We achieve this through (1) robust $k$-FWER control in the presence of complex dependencies, (2) improved reproducibility through reduced randomness, and (3) flexibility to accommodate mixed data types, including continuous and categorical covariates \citep{kormaksson2021sequential}. By enabling  reliable variable selection under $k$-FWER control, our approach significantly contributes to the design and interpretation of clinical trials and biomarker studies in precision medicine.
	
	
	\section{Methods}
	
	In high-dimensional variable selection problems, it is essential to control the inclusion of false discoveries to ensure the robustness and reliability of the results. While the False Discovery Rate (FDR) is a commonly used criterion, it focuses on controlling the expected proportion of false positives among the selected variables. However, in some applications, such as clinical research and genomic studies, even a small number of false discoveries can have significant consequences.
	
	The $k$-familywise error rate ($k$-FWER) provides a more stringent error control by limiting the probability of making at least \(k\) false discoveries. Formally, $k$-FWER is defined as:
	\[
	\text{$k$-FWER} = \Pr(V \geq k),
	\]
	where \(V\) is the number of false discoveries among the selected variables. When \(k = 1\), $k$-FWER reduces to the traditional familywise error rate (FWER), controlling the probability of making any false discoveries.
	
	In confirmatory studies, such as clinical trials or biomarker discovery, controlling $k$-FWER is particularly important because even a small number of false discoveries can lead to incorrect conclusions or wasted resources. For example, setting \( k = 2 \) permits up to one false discovery while controlling the probability of making two or more false discoveries. This balance between stringent error control and sufficient power makes $k$-FWER suitable for applications that require high reliability and reproducibility.
	
	\subsection{Knockoffs}
	The knockoff filter, originally developed by \citep{barber2015controlling}, is a powerful framework for variable selection with FDR control. This technique involves creating synthetic copies of the original variables, referred to as knockoffs, which act as negative controls when analyzing the response variable \( Y \). Once the knockoffs are generated, the knockoff filter identifies variables that demonstrate significantly greater importance than their knockoff counterparts, based on various measures of feature importance that can be calculated using several popular methods. We are going to provide the detailed information about every step.
	
	\textbf{Step 1: Construct Knockoffs.} 
	For each original variable \(X_j\), we create a corresponding artificial variable \(\tilde{X}_j\), known as a knockoff. These knockoff variables serve as controls to help assess the significance of the original variables. The knockoff variables \(\tilde{\mathbf{X}} = (\tilde{X}_1, \ldots, \tilde{X}_p)\) must satisfy two key properties:
	
	\begin{itemize}
		\item \textbf{Exchangeability:} The original variables and their knockoffs should be pairwise exchangeable, meaning that the joint distribution of \((\mathbf{X}, \tilde{\mathbf{X}})\) remains unchanged if we swap any original variable \(X_j\) with its knockoff \(\tilde{X}_j\).
		\item \textbf{Conditional Independence:} The knockoff variables should be conditionally independent of the outcome \(Y\) given the original variables \(\mathbf{X}\), i.e., \(\tilde{\mathbf{X}} \perp Y \mid \mathbf{X}\).
	\end{itemize}
	
	\textbf{Step 2: Calculate Knockoff Statistics for Penalized Cox Regression.}
	In this step, we use the LASSO method to solve the Cox regression and derive knockoff statistics based on the maximum lambda value at which each variable's coefficient becomes nonzero.
	
	First, we fit a penalized Cox regression model using the original and knockoff variables. The Cox proportional hazards model is commonly used for analyzing clinical trial data. The hazard function \(\lambda(t|\mathbf{X}_i)\) for subject \(i\) with covariates \(\mathbf{X}_i\) is given by:
	\begin{equation}
		\lambda(t|\mathbf{X}_i) = \lambda_0(t) \exp(\mathbf{X}_i^T \beta),
	\end{equation}
	where \(\lambda_0(t)\) is the baseline hazard function and \(\beta\) is the vector of regression coefficients.
	
	The partial likelihood function for the Cox model is given by:
	\begin{equation}
		L(\beta) = \sum_{i=1}^n \delta_i \left( \mathbf{X}_i^T \beta - \log \sum_{j:Y_j \geq Y_i} \exp(\mathbf{X}_j^T \beta) \right),
	\end{equation}
	where \(\delta_i\) is the event indicator for subject \(i\).
	
	The LASSO penalized Cox regression \citep{tibshirani1996regression} adds an \(L1\) penalty to the partial likelihood, encouraging sparsity in the model coefficients. The optimization problem is:
	\begin{equation}
		\hat{\beta} = \arg \min_{\beta} \left\{ -L(\beta) + \lambda \sum_{j=1}^p |\beta_j| \right\},
	\end{equation}
	where \(\lambda\) is a tuning parameter that controls the strength of the \(L1\) penalty.
	
	For each original variable \( X_j \), we determine the upper bound of \( \lambda \), at which its coefficient \( \hat{\beta}_j \) becomes nonzero. Formally, we define the statistic \( Z_j \) as:
	\begin{equation}
		\label{Z_stat}
		Z_j = \sup \{ \lambda : \hat{\beta}_j (\lambda) \neq 0 \}.
	\end{equation}
	
	Similarly, for each knockoff variable \( \tilde{X}_j \), we define the corresponding statistic \( \tilde{Z}_j \) analogously:
	\begin{equation*}
		\tilde{Z}_j = \sup \{ \lambda : \hat{\tilde{\beta}}_j (\lambda) \neq 0 \}.
	\end{equation*}
	
	The knockoff statistics \( W_j \) for each variable are then calculated as follows:
	\begin{equation}
		\label{knock_stat}
		W_j = \max \{ Z_j, \tilde{Z}_j \}, \quad \chi_j = \text{sgn}(Z_j - \tilde{Z}_j),
	\end{equation}
	where the sign function \(\text{sgn}(x)\) returns \(-1\), \(0\), or \(1\) depending on whether \( x < 0 \), \( x = 0 \), or \( x > 0 \), respectively.
	
	As discussed in prior literature, various statistical approaches can be employed, including least squares-based methods, which have been utilized to construct knockoff statistics for Cox regression, as demonstrated in \cite{hu2024model}. However, for datasets containing mixed data types, the \( Z \) statistics defined in Equation \ref{Z_stat} are advantageous because they do not require rescaling, which can introduce additional variability and controversy when applied to binary and continuous covariates.

	\textbf{Step 3: Calculate a data-dependent threshold for the statistics to control $k$-FWER.}
	
	By using these statistics, we can determine the importance of each original variable relative to its knockoff. The larger the value of \( W_j \), the more significant the variable is deemed to be in the context of the Cox model. This approach allows us to effectively control the FWER while identifying important variables in high-dimensional survival data. We extend this framework to control the $k$-FWER using the following knockoffs-based procedure as \citep{Janson2016} developed.
	
	The procedure for controlling the $k$-FWER is guided by comparing the importance statistics \( W_j \) and their associated signs \( \chi_j \) defined from Equation \ref{knock_stat}, where the sign indicates whether the knockoff or the original variable is more significant. A threshold \( T_v \) is chosen adaptively based on the observed data to ensure that the number of false discoveries is properly controlled. The key steps are outlined below:
	
	\begin{itemize}
		\item (i) Compute and order the importance statistics. \\
		Calculate the importance statistic \( W_j \) for each variable. Sort the statistics in decreasing order:
		\[
		W_{\rho(1)} \geq W_{\rho(2)} \geq \dots \geq W_{\rho(p)},
		\]
		where \( \rho(1), \rho(2), \dots, \rho(p) \) is a permutation of the indices \( \{1, 2, \dots, p\} \) that arranges the statistics from largest to smallest.
		
		\item (ii) 
		Traverse the ordered sequence of statistics \( W_{\rho(j)} \) and count the occurrences of negative signs \( \chi_j = -1 \). Let \( j^* \) be the index corresponding to the \( v \)-th occurrence of a negative sign in the sequence \( \chi_{\rho(1)}, \chi_{\rho(2)}, \dots, \chi_{\rho(p)} \). If fewer than \( v \) negative signs are observed, set \( j^* = p \).
		
		\item (iii)
		Set the threshold \( T_v \) to be:
		\[
		T_v = \sup \left\{ t > 0 : \# \{ j : W_j \geq t \ \text{and} \ \chi_j = -1 \} = v \right\}.
		\]
		If no such threshold exists, set \( T_v = -\infty \). This threshold ensures that at most \( v \) variables with negative signs are selected.
		
		\item (iv)
		Reject all null hypotheses \( H_{0,j} \) for variables satisfying:
		\[
		W_j \geq T_v \quad \text{and} \quad \chi_j = +1.
		\]
	\end{itemize}
	
	To determine the appropriate value of \( v \) for a given significance level \( \alpha \), we rely on the following theoretical result, which ensures that the $k$-FWER is properly controlled.

	\begin{prop}(Theorem 3.1 of \citet{Janson2016})
		Let \( V \) represent the number of false discoveries. For any integer \( k \geq 1 \) and significance level \( 0 < \alpha < 1 \), the parameter \( v \) is selected as the largest integer satisfying:
		\begin{equation}
			\sum_{i = k}^{\infty} 2^{-i - v} \binom{i + v - 1}{i} \leq \alpha.
		\end{equation}
		With this choice of \( v \), the $k$-FWER is controlled at level \( \alpha \), ensuring:
		\[
		\mathbb{P}(V \geq k) \leq \alpha.
		\]
	\end{prop}
	
	This procedure provides robust error control while enabling the identification of important variables in high-dimensional survival analysis.

	\subsection{Derandomized Knockoffs}
	In high-dimensional variable selection, the original knockoffs procedure has shown effectiveness in controlling error rates and identifying important variables. However, due to its reliance on the stochastic generation of knockoff copies, different runs of the algorithm can lead to varying selected variable sets, introducing instability in the selection process. This variability can result in uncertainty regarding the reliability of the selected features, especially in large-scale applications. To address this limitation, the derandomized knockoffs procedure \citep{ren2023derandomizing} aggregates selection outcomes from multiple independent runs of the knockoffs algorithm, thereby improving stability and enhancing the robustness of variable selection.
	
	The derandomized knockoffs method is designed to control the $k$-FWER by combining the results of multiple knockoff executions, as illustrated in Algorithm \ref{alg_derandom}. By averaging the selection frequencies across runs, this approach mitigates the variability inherent in individual knockoff runs and provides more reliable selection outcomes.
	
	\begin{algorithm}
		\caption{Derandomized Knockoffs Procedure (Algorithm 1 of \cite{ren2023derandomizing})}
		\begin{algorithmic}
			\State \textbf{Input:} Covariate matrix \(\mathbf{X} \in \mathbb{R}^{n \times p}\); response variables \(Y \in \mathbb{R}^n\); number of realizations \(M\); a base procedure; selection threshold \(\eta\).
			\State \textbf{1.} for \(m = 1, \ldots, M\) do
			\begin{itemize}
				\item[i.] Generate a knockoff copy \(\tilde{\mathbf{X}}^{(m)}\).
				\item[ii.] Run the base procedure with \(\tilde{\mathbf{X}}^{(m)}\) as knockoffs and obtain the selection set \(\hat{S}^{(m)}\).
			\end{itemize}
			\State end
			\State \textbf{2.} Calculate the selection probability:
			\[
			\hat{\pi}_j = \frac{1}{M} \sum_{m=1}^{M} 1\{j \in \hat{S}^{(m)}\}.
			\]
			\State \textbf{Output:} Selection set \(\hat{S} = \{j \in [p] : \hat{\pi}_j \geq \eta\}\).
		\end{algorithmic}
		\label{alg_derandom}
	\end{algorithm}
	
	The derandomized knockoffs procedure effectively controls the $k$-FWER by adjusting two key parameters: the selection threshold \( \eta \) and the number of knockoff copies \( M \). By appropriately tuning these parameters, the expected count of false discoveries can be kept within desired limits, minimizing the probability of making at least \( k \) false discoveries. As discussed in Section 4 of \cite{ren2023derandomizing}, this method ensures strong $k$-FWER control while maintaining statistical power, even in complex high-dimensional scenarios.
	
	Beyond its stability improvements, the derandomized approach offers theoretical guarantees for error control, making it a robust choice for high-dimensional survival data analysis. By incorporating this method, we enhance the reliability of variable selection while maintaining strict control over $k$ -FWER, thus improving the overall validity of our findings in the context of survival analysis.

	\subsection{Sequential Knockoffs for Mixed Data Types}
	
	In clinical and biomedical studies, it is common to encounter datasets containing a mixture of continuous and categorical variables. This is particularly prevalent in clinical trials, genomic studies, and other medical research where patient demographics, genetic variants, and clinical outcomes are often recorded in various formats. To effectively generate knockoffs in such settings, specialized methods are necessary to account for the different data types. 
	
	To address this need, we employ Algorithm \ref{alg_seqknock}, the sequential knockoffs approach proposed by \citep{kormaksson2021sequential}, which generates knockoffs sequentially while respecting the type of each variable. This method is incorporated in Step 1 of our procedure to ensure accurate knockoff generation across mixed data types. 
	
	\begin{algorithm}
		\caption{Sequential Knockoff Algorithm for Mixed Data Types (Algorithm 2 of \cite{kormaksson2021sequential})}
		\begin{algorithmic}[1]
			\State \textbf{Input:} Covariate matrix $\mathbf{X} \in \mathbb{R}^{n \times p}$ with mixed data types.
			\State \textbf{Step 1:} For each variable $X_j$, estimate the conditional distribution of $X_j$ given the other variables.
			\begin{itemize}
				\item \textbf{Continuous Variables:} If $X_j$ is continuous, fit a penalized linear regression model and sample the knockoff variable $\tilde{X}_j$ from the estimated conditional distribution:
				\[
				\tilde{X}_j \sim \mathcal{N}(\hat{\mu}_j, \hat{\sigma}_j^2),
				\]
				where $\hat{\mu}_j$ and $\hat{\sigma}_j^2$ denote the estimated mean and variance.
				\item \textbf{Categorical Variables:} If $X_j$ is categorical, fit a penalized multinomial logistic regression model and sample $\tilde{X}_j$ from the estimated multinomial distribution.
			\end{itemize}
			\State \textbf{Step 2:} Return the knockoff matrix $\tilde{\mathbf{X}} = (\tilde{X}_1, \ldots, \tilde{X}_p)$.
		\end{algorithmic}
		\label{alg_seqknock}
	\end{algorithm}
	
	By using sequential knockoffs, we effectively handle datasets with mixed data types, ensuring that knockoff variables are appropriately generated for both continuous and categorical features. This flexibility is essential in clinical data analysis, where covariates often originate from diverse sources, such as laboratory measurements, genetic data, and patient records. Furthermore, incorporating sequential knockoffs enables us to maintain rigorous control over the $k$-FWER while addressing the complexity of real-world data. This capability broadens the applicability of our method in high-impact fields, including medical research and personalized medicine.
	
	\subsection{Methods Comparison}
	\label{method_comp}
	We compare the Cox regression-version derandomized knockoffs procedure against several commonly used variable selection methods, including Lasso-based selection \citep{tibshirani1996regression}, the Benjamini-Hochberg (BH) procedure applied to p-values from Cox proportional hazards models \citep{benjamini1995controlling}, and stepwise regression based on the Akaike Information Criterion (AIC) \citep{hocking1976biometrics}. Additionally, we consider the vanilla knockoffs procedure based on Cox regression \citep{li2023coxknockoff,hu2024model}, which generates knockoffs without derandomization, and the derandomized knockoffs procedure based on linear regression \citep{ren2023derandomizing}. 
	
	\section{Simulations}
	To evaluate the performance of our method under realistic conditions, we design a simulation study with mixed data types, reflecting the complexity typically encountered in clinical and high-dimensional studies. 
	\subsection{Simulated Datasets Generation}
	
	We simulate the rows of the \( n \times p \) design matrix \( \mathbf{X} \) independently from a multivariate Gaussian distribution with mean \( \mathbf{0} \) and covariance matrix \( \Sigma = (\Sigma_{ij}) \). To cover a range of correlation structures, we consider two types of covariance matrices:
	\[
	\Sigma_{ij} = 
	\begin{cases} 
		I_{\{i=j\}}/n, & \text{(Independent)} \\ 
		\rho^{|i-j|}/n+, & \text{(AR1 structure)}.
	\end{cases}
	\]
	As \citep{kormaksson2021sequential} suggested, we fixed $\rho-0.5$ in this study.
	
	The design matrix \( \mathbf{X} \) consists of a mix of continuous and binary covariates. Continuous covariates are generated directly from the multivariate Gaussian distribution. To introduce binary covariates, we randomly select \( p_b \) of the \( p \) columns and dichotomize them using the indicator function \( \delta(x) = 1(x > 0) \). This setup reflects realistic applications where both continuous and categorical variables coexist, as seen in clinical and biomedical data.
	The survival times are then generated using an exponential distribution, defined as:
	\[
	T_i = -\frac{\log(U_i)}{\lambda \exp(\mathbf{X}_i \boldsymbol{\beta})},
	\]
	where \( U_i \sim \text{Uniform}(0, 1) \) is a random variable, \( \lambda \) is the baseline hazard rate, and \( \mathbf{X}_i \boldsymbol{\beta} \) represents the linear predictor for observation \( i \).
	
	Censoring times \( C_i \) are generated independently from a uniform distribution \( C_i \sim \text{Uniform}(0, 1) \). The observed survival time \( O_i \) and event indicator \( \delta_i \) for observation \( i \) are then defined as:
	\[
	O_i = \min(T_i, C_i), \quad \delta_i = 
	\begin{cases} 
		1 & \text{if } T_i \leq C_i \quad \text{(event occurred)} \\ 
		0 & \text{otherwise} \quad \text{(right-censored)}.
	\end{cases}
	\]
	
	The final dataset captures both the survival outcomes and the associated covariates, structured to reflect the relationships between predictors and event times. It includes the observed survival times, event indicators, and the covariate matrix \( \mathbf{X} \), providing a comprehensive setup for assessing the effectiveness of variable selection methods in high-dimensional survival analysis. By simulating the interaction between important covariates and the survival outcomes, this framework enables rigorous evaluation of model performance under realistic conditions with mixed data types.

	\subsection{Simulation Design}
	\label{subsec_simudesign}
	We conducted extensive simulations to evaluate the effectiveness of variable selection methods in varying configurations of effect sizes, correlation structures, and covariate dimensions. Our goal is to assess how well the methods identify important covariates while maintaining control over false discoveries at different levels of signal strength and correlation.
	
	
	For each simulation, we generate datasets with \( n = 300 \) observations, with varying numbers of covariates \( p \), nonnull covariates \( p_1 \), binary covariates \( p_b \), nonnull binary covariates \( p_{1b} \), continuous covariates \( p_c \), and nonnull continuous covariates \( p_{1c} \), as outlined in Table \ref{tab:simulation_params}. The regression coefficients \( \beta \) for the nonnull covariates are set to nonzero values, while the coefficients for null covariates are set to zero.
	
	\begin{table}[h]
		\centering
		\caption{Simulation parameter configurations}
		\label{tab:simulation_params}
		\begin{tabular}{|c|cc|cc|cc|}
			\hline
			\textbf{Setting} & \textbf{\( p \)} & \textbf{\( p_1 \)} & \textbf{\( p_b \)} & \textbf{\( p_{1b} \)} & \textbf{\( p_c \)} & \textbf{\( p_{1c} \)} \\ 
			\hline
			1 & 15  & 7   & 5  & 2  & 10 & 5  \\
			2 & 30  & 15  & 10 & 5  & 20 & 10 \\
			3 & 60  & 15  & 20 & 5  & 40 & 10 \\
			\hline
		\end{tabular}
	\end{table}
	
	The effect sizes of the nonnull covariates vary across simulations to represent different levels of signal strength. 
	We simulated effect sizes for binary covariates ($\beta_{\text{bin}}$) by emulating each value in the set of \{0.001, 0.1, 0.5, 1, 2, 3\} and for continuous covariates ($\beta_{\text{cont}}$) by emulating each value in the set of \{0.005, 2, 5, 10, 15, 25\} The baseline hazard rate in the survival time generation is fixed at \( \lambda = 0.1 \).
	
	For each parameter configuration, we perform 100 independent simulations to account for variability in the data and provide robust performance estimates. The derandomized knockoffs procedure is applied with \( M = 30 \) knockoff copies, and the selection threshold \( \eta \) is set to \(  0.8 \), chosen to closely match the value ($\eta = 0.81$) used in the \citep{ren2023derandomizing} original derandomized knockoffs paper. 
	To control the $k$-FWER at a significance level \( \alpha = 0.1 \), we evaluate the cases \( k = 2 \) and \( k = 3 \). 
	We do not consider the case $k = 1$ because controlling the $k$-FWER in this scenario tends to be overly stringent and impractical for most real-world applications. The results of the simulations are averaged to assess the overall effectiveness of the variable selection methods, measuring both their ability to detect true signals and their control over false discoveries.
	
	To evaluate the performance of variable selection methods, we use the True Positive Proportion (TPP) as key metrics, which quantifies the proportion of correctly identified important covariates. TPP and $k$-FWER allow us to assess the balance between selection accuracy and error control under various parameter configurations and correlation structures.

	\subsection{Results}
	
	We evaluate the performance of the proposed derandomized knockoffs procedure across different simulation settings, comparing it with the alternative variable selection methods described in Section \ref{method_comp}. The results are presented for two types of covariance structures: independent covariates and AR1-correlated covariates. The evaluation is based on two key metrics: the True Positive Proportion (TPP), which measures the proportion of true signals correctly identified, and the $k$-FWER, which is the primary focus of this study.
	
	\subsubsection{Performance under Independent Correlation Structure}
\begin{figure}[htbp]
		\center{\includegraphics[width=17cm]  {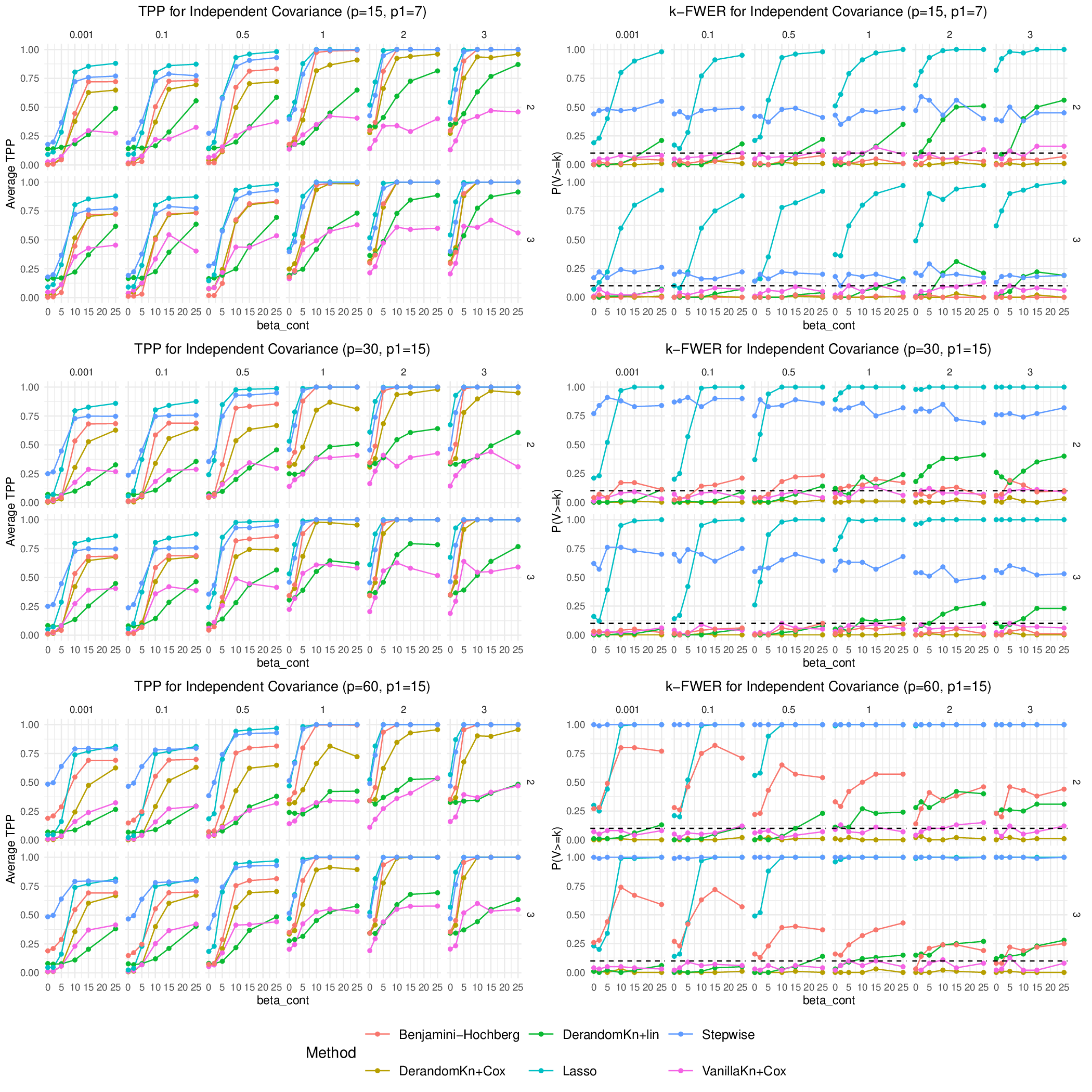}}
		\caption{Average selection performance of different variable selection methods under the independent covariance setting. The titles of each panel indicate the number of covariates ($p$) and nonnull features ($p_1$). The left panels display the True Positive Proportion (TPP) across varying signal strengths $\beta_{\text{cont}}$ and $\beta_{\text{bin}}$, while the right panels show the corresponding $k$-FWER. In each panel, the X-axis represents the values of $\beta_{\text{cont}}$, and the values marked at the top of each column denote the corresponding $\beta_{\text{bin}}$ values. The numbers $2$ and $3$ on the right of each row indicate the $k$ values for $k$-FWER. The dashed horizontal lines in the right panels represent the pre-specified significance level $\alpha = 0.1$ for $k$-FWER control.
		}
		\label{simu_inde}
	\end{figure}

	Figure~\ref{simu_inde} presents the results for the independent covariance setting. Across all methods, the TPP increases as the signal strength $\beta_{\text{cont}}$ and $\beta_{\text{bin}}$ grows. The derandomized knockoffs procedure based on Cox regression consistently achieves high TPP (above $80\%$ in most cases) while maintaining strict control over the $k$-FWER, whose $k$-FWER curves are always under proposed level $\alpha=0.1$ as the dashed horizontal lines show in the right panels. 
	
	In contrast, the vanilla knockoffs based on Cox regression and the derandomized knockoffs based on linear regression tend to have a higher $k$-FWER, which in some cases exceeds the pre-specified significance level, indicating weaker error control. Additionally, both methods exhibit lower TPP in most cases, particularly for weaker signals, further highlighting the advantages of aggregation in the derandomization process. The derandomized knockoffs based on Cox regression not only achieves stronger control of $k$-FWER but also consistently outperforms these alternatives in terms of TPP across different simulation settings.
	
	Among other competing methods, the Benjamini-Hochberg (BH) procedure also fails to maintain strong $k$-FWER control in some settings, particularly for higher number of features. Lasso-based selection  achieves competitive TPP but struggles to control the $k$-FWER, especially as the number of nonnull covariates increases.  Stepwise regression exhibits relatively low power across all settings. 
	
	\subsubsection{Performance under AR1 Correlation Structure}
\begin{figure}[htbp]
		\center{\includegraphics[width=17cm]  {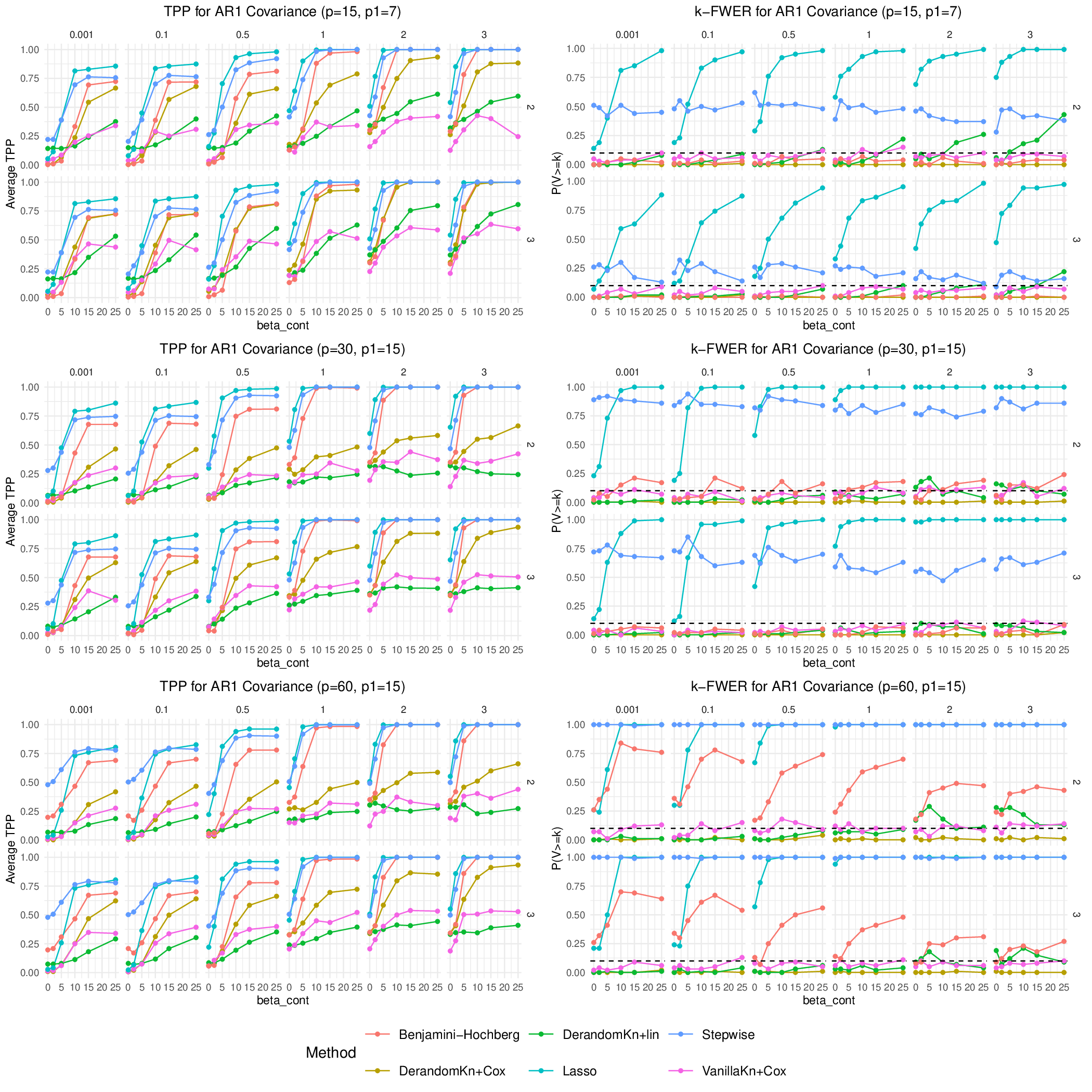}}
		\caption{Average selection performance of different variable selection methods under the AR1 covariance setting. }
		\label{simu_ar1}
	\end{figure}

	Figure~\ref{simu_ar1} presents the results for the AR(1) covariance setting, where covariates exhibit autocorrelation, making variable selection more challenging due to multicollinearity. As expected, TPP increases as the signal strength $\beta_{\text{cont}}$ and $\beta_{\text{bin}}$ grow. The derandomized knockoffs procedure based on Cox regression maintains strong control of the $k$-FWER, with its curves consistently remaining below the significance level $\alpha = 0.1$. However, in contrast to the independent case, its TPP is notably lower when $k = 2$, particularly for weaker signals, indicating that the added correlation among covariates reduces selection power. Nonetheless, for $k = 3$, the derandomized knockoffs procedure remains competitive, achieving a favorable trade-off between power and error control.
	
	Compared to the independent setting, all methods experience a reduction in TPP due to the increased difficulty of distinguishing correlated covariates. Despite this, the derandomized knockoffs method still demonstrates a clear advantage over the vanilla knockoffs based on Cox regression and the derandomized knockoffs based on linear regression, both of which tend to have higher $k$-FWER and lower TPP. The vanilla knockoffs approach, in particular, frequently exceeds the pre-specified significance level, further emphasizing the importance of aggregation in the derandomization process.
	
	Among other competing methods, the Benjamini-Hochberg (BH) procedure exhibits greater variability in its error control compared to the independent case, struggling to maintain $k$-FWER control as the number of features increases. Lasso-based selection continues to achieve competitive TPP but fails to control $k$-FWER, particularly in high-dimensional settings. Stepwise regression remains conservative, showing consistently low power across all configurations.
	
	Overall, while all methods show some loss of power in the AR1 setting due to the presence of correlation among covariates, the derandomized knockoffs procedure based on Cox regression remains the most reliable in balancing power and error control. Its performance gap with alternative methods is more pronounced in this more challenging scenario, reinforcing its robustness in structured high-dimensional data.
	
	\section{Clinical Data Application}
	
	
	To evaluate the performance of our method in a real-world setting, we apply it to clinical data from a study on primary biliary cirrhosis (PBC) \citep{fleming2013counting}, a chronic liver disease that can lead to liver failure. This dataset originates from a well-documented longitudinal study conducted at the Mayo Clinic between 1974 and 1984, which investigated risk factors affecting patient survival. The PBC dataset is widely used in survival analysis due to its high-dimensional structure and the presence of both continuous and categorical predictors, making it an ideal test case for our method. By applying our variable selection framework to this dataset, we aim to identify critical clinical factors associated with survival while maintaining strict control over false discoveries. After removing records with missing data and patients who received transplants, the sample contains $n = 258$ observations. The dataset originally includes 17 clinical and biochemical variables, with a mix of continuous variables such as age, serum bilirubin, albumin levels, and prothrombin time, along with categorical variables like sex, presence of ascites, and histologic stage of disease. 
	
\begin{figure}[htbp]
		\center{\includegraphics[width=17cm]  {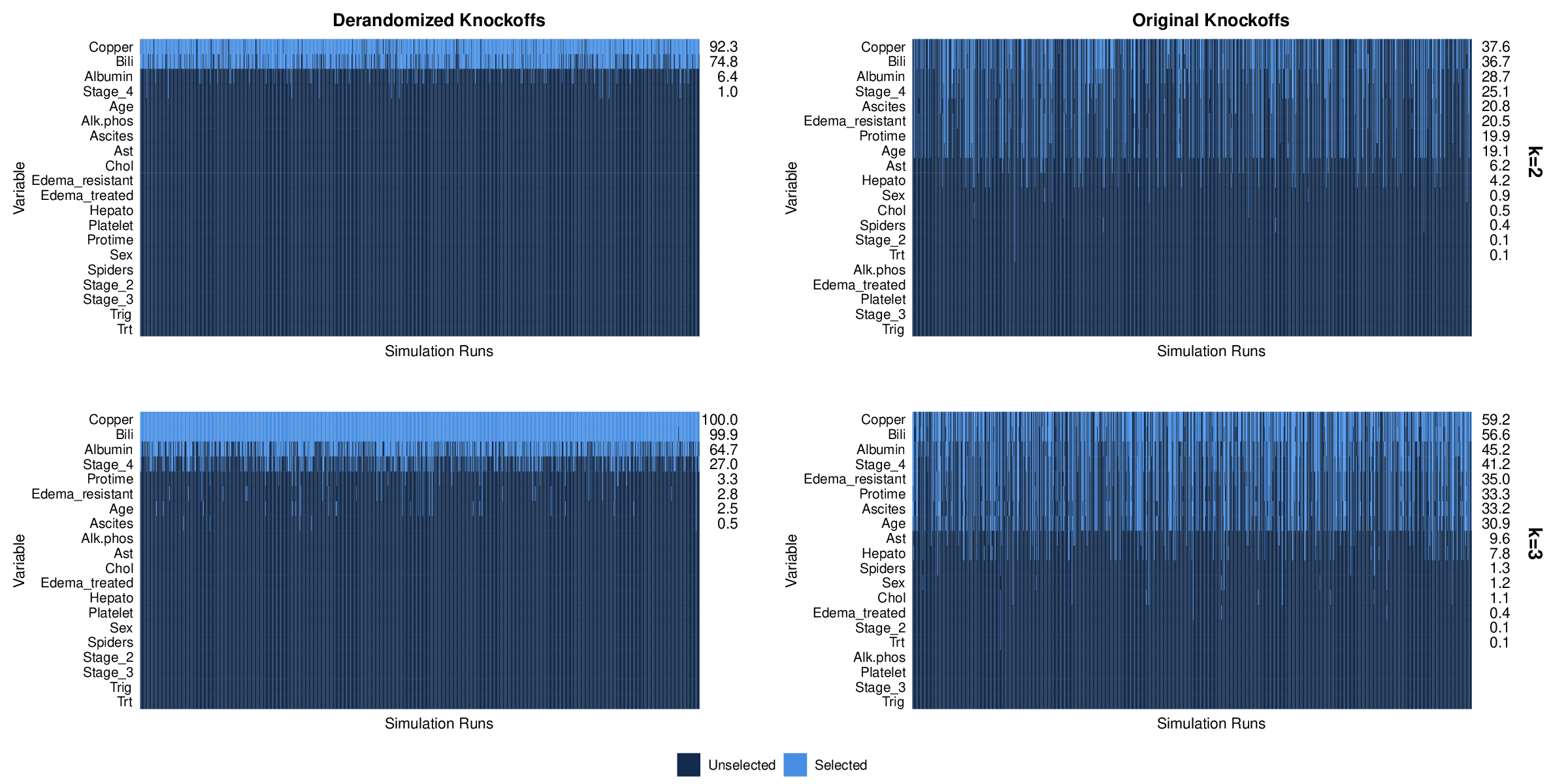}}
		\caption{Selection frequency heatmaps for the PBC dataset under different methods and $k$-FWER control levels. Each panel displays the selection results across 1000 simulation runs. The left column corresponds to the derandomized knockoffs procedure, while the right column corresponds to the original knockoffs procedure. The top row shows results for $k = 2$, and the bottom row for $k = 3$. Variables are ordered by their selection frequency in the derandomized knockoffs method (top-left panel), with the selection proportion (in percentage) displayed to the right of each heatmap. Dark blue indicates unselected instances, while light blue indicates selected instances. }
		\label{fig:pbc_summ}
	\end{figure}

	Since \textit{edema} and \textit{histologic stage of disease} have multiple categories, we transformed them into separate binary indicators for better interpretability. \textit{Edema}, originally coded as 0 (no edema), 0.5 (untreated or successfully treated edema), and 1 (edema despite diuretic therapy), was split into two binary variables: one indicating successfully treated edema and another indicating treatment-resistant edema. Similarly, \textit{histologic stage of disease}, which ranges from 1 to 4, was converted into three binary indicators for stages 2, 3, and 4. These modifications increased the total number of features from 17 to 20. Table~\ref{tab:pbc_variables} provides a detailed description of all variables used in our analysis, distinguishing between binary and continuous features. These variables will be examined in Figure~\ref{fig:pbc_summ} to assess their selection frequencies across different methods.
	
	\begin{table}[h]
		\centering
		\caption{Description of variables in the PBC dataset}
		\label{tab:pbc_variables}
		\begin{tabular}{|l|l|l|l|}
			\hline
			\multicolumn{2}{|c|}{\textbf{Binary Variables}} & \multicolumn{2}{c|}{\textbf{Continuous Variables}} \\ 
			\hline
			\textbf{Variable} & \textbf{Description} & \textbf{Variable} & \textbf{Description} \\ 
			\hline
			\textit{Ascites} & Presence of ascites & \textit{Age} & Age in years  \\ 
			\textit{Edema-resistant} & Indicator for treatment-resistant edema & \textit{Albumin} & Serum albumin (g/dL)  \\ 
			\textit{Edema-treated} & Indicator for successfully treated edema & \textit{Alk-phos} & Alkaline phosphatase (U/L)  \\ 
			\textit{Hepato} & Presence of hepatomegaly & \textit{Ast} & Aspartate aminotransferase (U/L)  \\ 
			\textit{Sex} & Gender (1/0 for male/female) & \textit{Bili} & Serum bilirubin (mg/dL)  \\ 
			\textit{Spiders} & Presence of spider angiomas & \textit{Chol} & Serum cholesterol (mg/dL)  \\ 
			\textit{Stage-2} & Histologic disease stage 2 & \textit{Copper} & Urine copper (ug/day)  \\ 
			\textit{Stage-3} & Histologic disease stage 3 & \textit{Platelet} & Platelet count (10$^3$ per mm$^3$)  \\ 
			\textit{Stage-4} & Histologic disease stage 4 & \textit{Protime} & Prothrombin time (seconds)  \\ 
			\textit{Trt} & \makecell[l]{Treatment group indicator \\ (1/0 for placebo/D-penicillamin)} & \textit{Trig} & Serum triglycerides (mg/dL)  \\ 
			\hline
		\end{tabular}
	\end{table}
	
	We applied the derandomized knockoffs procedure and the original knockoffs method, both based on Cox regression, to the primary biliary cirrhosis (PBC) dataset to evaluate their performance in real-world survival analysis. For derandomized knockoffs, similar to the settings described in Section~\ref{subsec_simudesign}, we set $\alpha=0.1$, $\eta=0.8$, and $M=30$. Figure~\ref{fig:pbc_summ} presents the selection frequency of variables across 1000 simulation runs under different values of $k$ in the $k$-FWER control framework ($k = 2$ in the top row and $k = 3$ in the bottom row). Variables are ranked based on their selection frequency in the derandomized knockoffs method (top-left panel), with selection proportions displayed on the right of each heatmap.
	
	Across different settings, the derandomized knockoff method demonstrated significantly better stability in identifying important variables. When controlling the $k$-FWER at $k=2$, the original knockoff method \citep{hu2024model} exhibited highly unstable selection results—all variables had selection probabilities below 40\%, and the selection frequencies decreased gradually without a clear threshold for identifying key predictors. In contrast, the derandomized knockoff method consistently selected Copper (92.3\%) and Bilirubin (74.8\%) with high probabilities, indicating a much clearer distinction between important and unimportant features.
	
	At $k=3$, where the error control is more relaxed, the instability of the original knockoff method remained evident. Only two variables (Copper and Bilirubin) had selection probabilities exceeding 50\%, while six others ranged between 30\% and 46\%, lacking a clear separation from the null features. Meanwhile, the derandomized knockoff method maintained a distinct selection pattern, with four variables surpassing 50\% selection frequency: Copper (100.0\%), Bilirubin (99.9\%), Albumin (64.7\%), and Stage-4 (27.0\%). All other variables had selection probabilities below $5\%$, making them easily distinguishable as null features.
	
	These results highlight the superior stability of the derandomized knockoff method in variable selection. Unlike the original knockoff method, which struggled to establish a clear importance threshold, our approach consistently ranked key prognostic markers with higher selection probabilities, reinforcing its effectiveness in identifying robust survival-associated biomarkers. The improved stability of the derandomized knockoff method also provides a practical advantage: it enables reliable feature selection with significantly fewer repetitions of the algorithm, reducing computational cost while maintaining accuracy in identifying important variables.
	
	\section{Conclusion}
	In this study, we adapt the derandomized knockoff procedure \citep{ren2023derandomizing} to Cox regression for variable selection in the high-dimensional survival analysis. By integrating the sequential knockoff framework \citep{kormaksson2021sequential} to handle mixed data types, which commonly arise in survival datasets, our method effectively identifies important biomarkers while rigorously controlling the $k$-FWER. Through extensive simulation studies, we demonstrated that the derandomized knockoffs approach consistently outperforms the original knockoffs and other conventional methods, exhibiting superior selection stability and clearer separation between true and null features. Our results highlight that the aggregation of knockoffs improves power and robustness, particularly in challenging settings with high-dimensional and correlated covariates.
	
	We further validated our method using the primary biliary cirrhosis (PBC) dataset, demonstrating its capability to select clinically relevant variables with high confidence. Compared to the original knockoffs approach, our method provides a clearer distinction between significant and nonsignificant features, offering a more distinct threshold to differentiate null from non-null variables. This facilitates a more reliable identification of prognostic biomarkers. Additionally, the enhanced stability of derandomized knockoffs suggests that fewer replications may be sufficient in practice, reducing computational demands without sacrificing selection accuracy.

	Despite these advantages, our approach relies on the assumption that the knockoff construction appropriately captures dependencies in the covariates. Future research could explore adaptive knockoff generation methods, such as deep knockoffs \citep{Romano2020}, which leverage machine learning techniques to better model complex covariate dependencies, potentially further enhancing power and applicability. 
	Additionally, while our current study focuses specifically on the Cox proportional hazards model, extensions to handle more complicated survival scenarios, including competing risks \citep{fine1999proportional} or time-varying covariates \citep{therneau2000cox}. Overall, our findings underscore the effectiveness of adapting derandomized knockoffs in high-dimensional biomarker selection, offering a powerful tool for statistical inference in clinical trial studies.

	\section*{Data Availability}
	
	The primary biliary cirrhosis (PBC) dataset used in this study is publicly available as part of the \texttt{survival} package in R. It originates from a well-documented clinical trial conducted at the Mayo Clinic between 1974 and 1984, investigating risk factors affecting patient survival \citep{fleming2013counting}. The dataset can be accessed and loaded directly in R using the following command:
	\begin{verbatim}
		library(survival)
		data(pbc)
	\end{verbatim}
	Additional details about the dataset and its variables can be found in the official documentation of the \texttt{survival} package.
	
	\section*{Code Availability}
	
	The code used in this study, including the implementation of the derandomized knockoff procedure for Cox regression and mixed-type data analysis, is publicly available on GitHub at:
	\url{https://github.com/jerryliu01998/DerandomKnock_Cox}
	This repository contains all scripts for data preprocessing, simulations, and real data analysis.
	
	\section*{Acknowledgment}
	
	This research was partially supported by the Bristol Myers Squibb Biostatistics Summer Internship Program.  
	
	\bibliographystyle{plainnat}
	\bibliography{Ref_knockoff}
	
\end{document}